\documentclass[prl,twocolumn,superscriptaddress]{revtex4-1}
\usepackage{graphicx}
\usepackage{amsmath,amssymb}
\usepackage[colorlinks=true,
            urlcolor=blue,
            linkcolor=blue,
            anchorcolor=blue,
            citecolor=blue]{hyperref}
\renewcommand{\section}[1]{{\par\it #1.---}\ignorespaces}
\begin{document}
\title{Retrieving ideal precision in noisy quantum optical metrology}
\author{Kai Bai}
\affiliation{School of Physical Science and Technology \& Key Laboratory for Magnetism and Magnetic Materials of the MoE, Lanzhou University, Lanzhou 730000, China}
\author{Zhen Peng}
\affiliation{School of Physical Science and Technology \& Key Laboratory for Magnetism and Magnetic Materials of the MoE, Lanzhou University, Lanzhou 730000, China}
\author{Hong-Gang Luo}
\affiliation{School of Physical Science and Technology \& Key Laboratory for Magnetism and Magnetic Materials of the MoE, Lanzhou University, Lanzhou 730000, China}
\affiliation{Beijing Computational Science Research Center, Beijing 100084, China}
\author{Jun-Hong An}
\email{anjhong@lzu.edu.cn}
\affiliation{School of Physical Science and Technology \& Key Laboratory for Magnetism and Magnetic Materials of the MoE, Lanzhou University, Lanzhou 730000, China}

\begin{abstract}
Quantum metrology employs quantum effects to attain a measurement precision surpassing the limit achievable in classical physics. However, it was previously found that the precision returns the shot-noise limit (SNL) from the ideal Zeno limit (ZL) due to the photon loss in quantum metrology based on Mech-Zehnder interferometer. Here, we find that not only the SNL can be beaten, but also the ZL can be asymptotically recovered in long-encoding-time condition when the photon dissipation is exactly studied in its inherent non-Markovian manner. Our analysis reveals that it is due to the formation of a bound state of the photonic system and its dissipative noise. Highlighting the microscopic mechanism of the dissipative noise on the quantum optical metrology, our result supplies a guideline to realize the ultrasensitive measurement in practice by forming the bound state in the setting of reservoir engineering.
\end{abstract}

\maketitle
\section{Introduction}\label{introduction}
Pursuing high-precision measurement to physical quantities, metrology plays a significant role in advancing the innovation of science and technology. Restricted by the unavoidable errors, the metrology precision realized in classical physics is strongly bounded by the shot-noise limit (SNL) $N^{-1/2}$ with $N$ being the number of resource employed in the measurements. It was found that the SNL can be beaten by taking advantage of the quantum effects such as squeezing \cite{PhysRevD.23.1693,MA201189,PhysRevLett.118.140401} and entanglement \cite{Nagata726,PhysRevLett.112.103604,Luo620}. This inspires the birth of a newly emerged field, quantum metrology \cite{Giovannetti1330,PhysRevLett.96.010401,RevModPhys.90.035005}. Many fascinating applications of quantum metrology have been proposed. The quantum effects of light can offer enhanced imaging resolution \cite{1464-4266-4-3-372,PhysRevX.6.031033,PhysRevLett.117.190802} in biological monitoring \cite{doi:10.1063/1.4724105,PhysRevX.4.011017,Taylor20161} and in optical lithography \cite{PhysRevLett.85.2733}, and improved sensitivity in gravitational wave detection \cite{PhysRevLett.110.181101} and in radar \cite{PhysRevLett.114.080503}. The quantum characters of atoms or spins can provide an enhanced precision in sensing weak magnetic field \cite{Jones1166,PhysRevLett.112.150801,PhysRevLett.113.103004,11,RevModPhys.89.035002,Chalopin} and ultimate accuracy for clocks \cite{YeJ2014,PhysRevLett.117.143004,Hosten}.

\begin{figure}[tbp]
	\centering
	\includegraphics[height=.43\columnwidth]{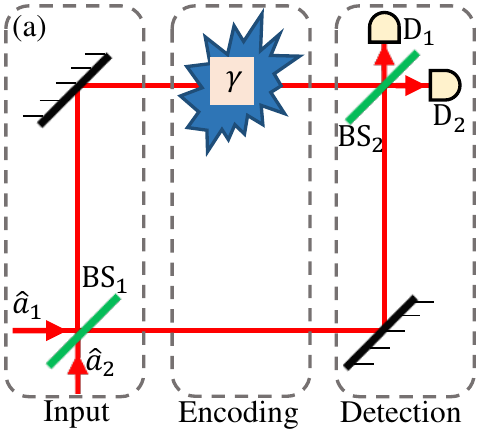}~~\includegraphics[height=.45\columnwidth]{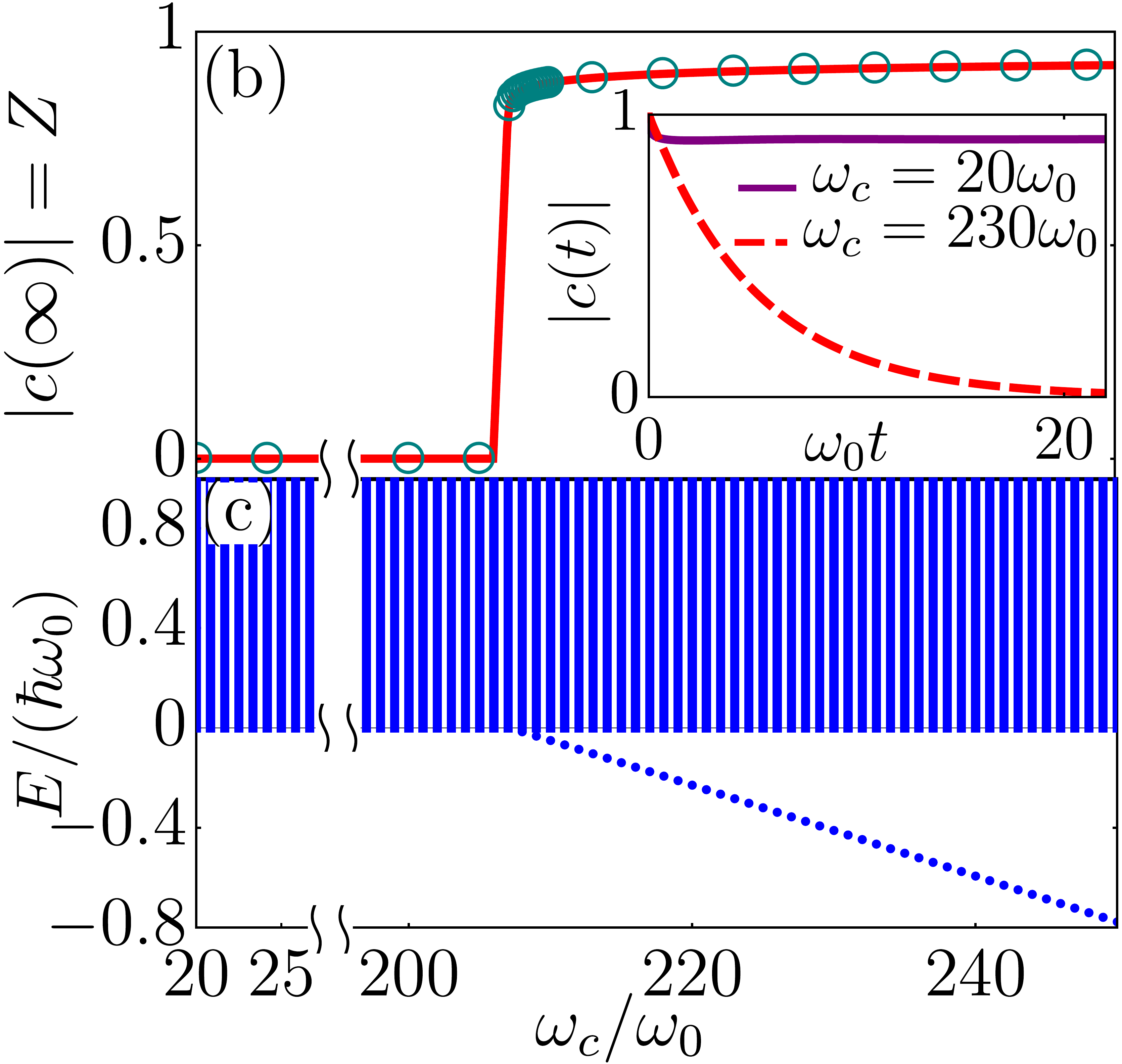}\\
	\caption {(a) Scheme of MZI-based quantum metrology. Two fields interact at beamsplitter $\text{BS}_1$ and propagate along two arms. One of the fields couples to a system with the potential influence of quantum noise, by which the estimated parameter $\gamma$ is encoded. After interfering at $\text{BS}_2$, the fields are detected by the detectors $\text{D}_1$ and $\text{D}_2$. (b) Long-time behavior of $|c(t)|$ (dark cyan circles) by solving Eq. \eqref{eq:8}, which coincides with $Z$ (red solid line) from the bound-state analysis. The inset shows the evolution of $|c(t)|$. (c) Energy spectrum of the whole system of the optical field and its environment. The parameters $s=1$, $\gamma=\pi\omega_0$, and $\eta=0.02$ are used. }
	\label{MZ}
\end{figure}

A wide class of quantum metrology using quantized light as probe is generally based on the Mach-Zehnder interferometer (MZI). Caves first pointed out that the precision can beat the SNL and reach the scale as $N^{-3/4}$ with the mean photon number $N$ by using the squeezed light \cite{PhysRevD.23.1693,HLL}. The scaling is named as Zeno limit (ZL) \cite{PhysRevLett.109.233601}.
Since then, a host of quantum states of light, such as N00N state \cite{PhysRevLett.85.2733}, twin Fock state \cite{PhysRevLett.71.1355}, two-mode squeezed state \cite{PhysRevLett.104.103602}, and entangled coherent state \cite{PhysRevLett.107.083601,PhysRevA.88.043832}, have been found to perform well in quantum optical metrology.
However, the decoherence caused by the unavoidable interactions of the optical probe with its environment degrades the real performance of quantum metrology \cite{PhysRevA.81.033819,PhysRevA.90.033846,PhysRevLett.102.040403,PhysRevA.81.033819,PhysRevA.80.013825,PhysRevA.95.053837,PhysRevA.78.063828,Gilbert:08,PhysRevA.75.053805,ZhangWP,ysw}, which hinders its practical application \cite{Banaszek2009}. It was found that the N00N state losses its metrological advantage when even a single photon is absorbed by the environment \cite{PhysRevLett.102.040403}. Entangled coherent state can beat the SNL, but only when the photon loss is extremely small \cite{PhysRevLett.107.083601,PhysRevA.88.043832}. This was proven to be true for the squeezed \cite{PhysRevA.81.033819,PhysRevA.95.053837} and definite-photon-number \cite{PhysRevLett.102.040403,PhysRevA.80.013825} states. When the photon loss is severe, the precision returns the SNL. It is called no-go theorem for noisy quantum metrology \cite{Al2018}. In those works, the decoherence of the probe was phenomenologically described by a transmissivity, which is equivalent to a continuous photon loss governed by a Born-Markovian master equation with constant loss rate \cite{PhysRevLett.102.040403,PhysRevLett.108.130402,Lu2015}. Given the inherent non-Markovian character of the decoherence dynamics \cite{PhysRevA.76.042127,PhysRevA.81.052330,PhysRevLett.109.170402,PhysRevLett.121.220403,Strathearn2018,RevModPhys.88.021002,Rivas_2014,LI20181}, it is expected that such treatment is insufficient. It was really found in the Ramsey-spectroscopy-based quantum metrology that the non-Markovian effect can transiently make the precision surpass the SNL in dephasing noises with Ohmic spactral density \cite{PhysRevA.84.012103,PhysRevLett.109.233601,PhysRevA.92.010102}, which is in sharp contrast to the Markovian approximate result \cite{PhysRevLett.79.3865}. However, the precision gets worse and worse in long-encoding-time condition.

In this Letter, we presents a physical mechanism to overcome this ostensible no-go theorem via the exact decoherence description to the optical probe in MZI-based quantum metrology. Focusing on the long-encoding-time condition, which corresponds to the case of full loss of photon and the quantum superiority completely disappears in the phenomenological description \cite{PhysRevLett.102.040403,PhysRevLett.108.130402,Lu2015}, we show that the ideal ZL is asymptotically recovered when a coherent state in one input port and squeezed state in the other are employed. Our analysis reveals that such recovery to the ideal precision is intrinsically caused by the formation of a bound state in the energy spectrum of the probe and its environment. On the one hand our result demonstrates that the phenomenological treatment overestimates the detrimental influences of the decoherence on quantum optical metrology, and on the other hand it supplies a guideline to realize the ultrasensitive measurement in practice by forming the bound state in the setting of reservoir engineering.

\section{Ideal quantum metrology}
To estimate a parameter of a system, one generally prepares a probe and couples it to the system to encode the parameter information. Then a series of measurements to certain observable are made to the probe. The value and the uncertainty of the parameter are estimated from the measurement results \cite{PhysRevLett.96.010401}. Consider the estimation of a frequency parameter $\gamma$ of the system. We choose two modes of optical fields with frequency $\omega_0$ as the probe. The encoding of $\gamma$ is realized by the time evolution $\hat{U}_0(\gamma,t)=\exp(-{i\hat{H}_0t/\hbar})$ with
\begin{equation}
\hat{H}_0=\hbar\omega_0\sum_{m=1,2}\hat{a}_m^\dag\hat{a}_m+\hbar\gamma\hat{a}_2^\dag\hat{a}_2,
\end{equation}where the first term is the Hamiltonian of the fields and the second one is the linear interaction of the second field with the system \cite{PhysRevLett.107.083601,PhysRevLett.101.040403,PhysRevA.77.012317}. The evolution $\hat{U}_0(\gamma,t)$ accumulates a phase difference $\gamma t$ to the two fields, which is measured by the MZI. The MZI has two beam splitters $\text{BS}_i$ ($i=1,2$) separated by the phase shifter $\hat{U}_0(\gamma,t)$ and two detectors $\text{D}_i$ [see Fig. \ref{MZ}(a)] \cite{DEMKOWICZDOBRZANSKI2015345}. Its input-output relation reads
$|\Psi_\text{out}\rangle=\hat{V}\hat{U}_0(\gamma,t)\hat{V}|\Psi_\text{in}\rangle$, where $\hat{V}=\exp[i\frac{\pi}{4}(\hat{a}_{1}^{\dagger}\hat{a}_{2}+\hat{a}_{2}^{\dagger}\hat{a}_{1})]$ is the action of $\text{BS}_i$ \cite{PhysRevLett.104.103602}. We are interested in the quantum superiority of $|\Psi_\text{in}\rangle$ in metrology subject to explicit measurement scheme. Thus we consider Caves's original scheme \cite{PhysRevD.23.1693}, i.e., the photon difference $\hat{M}=\hat{a}^\dag_1\hat{a}_1-\hat{a}^\dag_2\hat{a}_2$ is measured by $\text{D}_i$, which is also the most general measurement in MZI. Note although it does not saturate the Cram\'{e}r-Rao bound governed by quantum Fisher information \cite{PhysRevLett.100.073601}, it sufficiently demonstrate the quantum superiority especially in an experimentally friendly manner.

We consider $|\Psi_\text{in}\rangle=\hat{D}_{\hat{a}_{1}}\hat{S}_{\hat{a}_{2}}|0,0\rangle$, where $\hat{D}_{\hat{a}}=\exp(\alpha \hat{a}^{\dagger}-\alpha^{*}\hat{a})$ with $\alpha=|\alpha|e^{i\varphi}$, $\hat{S}_{\hat{a}}=\exp[\frac{1}{2}(\xi^{*}\hat{a}^{2}-\xi \hat{a}^{\dagger2})]$ with $\xi=re^{i\phi}$, and $|0,0\rangle$ is the two-mode vacuum state. Its total photon number is $N=|\alpha|^2+\sinh^2r$, which contains the ratio $\beta\equiv\sinh^2r/N$ from the squeezed mode and can be regarded as the quantum resource of the scheme. To the output state $|\Psi_\text{out}\rangle$ we can calculate $\bar{M} =[\sinh ^{2}r-\left\vert \alpha \right\vert ^{2}]\cos \gamma t$ and
\begin{eqnarray}
  \delta M &=& \{\cos ^{2}\gamma t[\left\vert \alpha \right\vert ^{2}+2\sinh^{2}r\cosh ^{2}r]+\sin ^{2}\gamma t\nonumber\\
  &&\times[\left\vert \alpha \cosh r-\alpha ^{\ast}\sinh re^{i\phi }\right\vert ^{2}+\sinh ^{2}r]\}^{1\over2},
\end{eqnarray}
where $\delta M=[\overline{M^2}-\bar{M}^2]^{1/2}$ and $\bar{\bullet}=\langle\Psi_\text{out}|\hat{\bullet}|\Psi_\text{out}\rangle$. They characterize the statistical distribution of the measurement results to $\hat{M}$. Then the best precision of estimating $\gamma$ can be evaluated by $\delta \gamma=\frac{\delta M}{|\partial\bar{M}/\partial\gamma|}$ as
\begin{equation}\label{eq:2}
\min\delta \gamma =\frac{[(1-\beta)e^{-2r}+\beta]^{1\over2}}{t\sqrt{N}|1-2\beta|},
\end{equation}
when $\phi =2\varphi $ and $\gamma t={(2m+1)\pi /2}$ for $m\in \mathbb{Z}$. If the squeezing is absent, then $\min\delta \gamma|_{\beta=0}=(t N_0^{1/2})^{-1}$ with $N_0=|\alpha|^2$ is just the SNL. For $\beta\neq 0$, using $e^{-2r}\simeq 1/(4\sinh^2r)$ for $N\gg1$ and optimizing $\beta$, we obtain
\begin{equation}\label{znl}
  \min\delta\gamma|_{\beta=(2\sqrt{N})^{-1}}=(tN^{3/4})^{-1},
\end{equation}
which is the ZL. It beats the SNL and manifests the superiority of the squeezing in metrology \cite{PhysRevD.23.1693,DEMKOWICZDOBRZANSKI2015345}. Benefited from the quantum-enhanced sensitivity, the squeezing has been used in gravitational-wave observatory \cite{PhysRevLett.110.181101}.
\section{Effects of dissipative noises} \label{effects}
In reality, the decoherence caused by the unavoidable interaction of the probe with the environment would deteriorate the performance of the quantum metrology. Conventionally, the decoherence of the optical probe is phenomenologically analyzed by introducing a transmissivity \cite{PhysRevA.81.033819,PhysRevA.90.033846,PhysRevLett.102.040403,PhysRevA.81.033819,PhysRevA.80.013825,PhysRevA.95.053837,PhysRevA.78.063828,Gilbert:08,PhysRevA.75.053805,ZhangWP}. This is equivalent to a continuous photon loss described by a Born-Markovian master equation \cite{PhysRevLett.102.040403,PhysRevLett.108.130402,Lu2015}. Recently, people found that the system-environment interplay caused by the inherent non-Markovian nature would induce diverse characters absent in the Born-Markovian approximate treatment \cite{PhysRevA.76.042127,PhysRevA.81.052330,PhysRevLett.109.170402,PhysRevLett.121.220403,Strathearn2018,RevModPhys.88.021002,Rivas_2014,LI20181}. To uncover the constructive role played by the non-Markovian effect in quantum metrology, we here investigate the exact decoherence dynamics of the optical probe and evaluate its metrology scale, especially in the long-encoding-time condition.

Taking the dissipative noise into account, the Hamiltonian governing the parameter encoding reads
\begin{equation}
\hat{H}=\hat{H}_0+\hbar\sum_k\omega_k[\hat{b}_k^\dag\hat{b}_k+g_k(\hat{a}_2\hat{b}_k^\dag+\text{H.c.})],
\end{equation}
where $\hat{b}_{k}$ is the annihilation operators of the $k$th environmental mode with frequency $\omega_{k}$ and $g_{k}$ is its coupling strength to the probe. The coupling is characterized by the spectral density $J(\omega)=\sum_kg_k^2\delta(\omega-\omega_k)$. In the continuum limit, it reads $J(\omega)=\eta\omega(\frac{\omega}{\omega_{c}})^{s-1}e^{-\frac{\omega}{\omega_{c}}}$, where $\eta$ is a coupling constant, $\omega_{c}$ is a cutoff frequency, and the exponent $s$ classifies the noise into sub-Ohmic for $0<s<1$, Ohmic for $s=1$, and super-Ohmic for $s>1$ \cite{RevModPhys.59.1}.

In the Heisenberg picture with $\hat{\bullet}(t)=\hat{U}^\dag(\gamma,t)\hat{\bullet}\hat{U}(\gamma,t)$ and $\hat{U}(\gamma,t)=\exp({-i\hat{H}t/\hbar})$, we can calculate $\hat{a}_1(t)=\hat{a}_1\exp(-i\omega_0t)$ and $\hat{a}_{2}(t)=c(t)\hat{a}_{2}+\sum_{k}d_{k}(t)\hat{b}_{k}$ \cite{PhysRevA.81.052105} with $|c(t)|^{2}+\sum_k|d_{k}(t)|^{2}=1$ satisfying
\begin{eqnarray}\label{eq:8}
\dot{c}(t)+i(\gamma+\omega_{0})c(t)+\int_{0}^{t}f(t-\tau)c(\tau)d\tau=0,
\end{eqnarray}where $f(t-\tau)=\int_0^\infty J(\omega)e^{-i\omega(t-\tau)}$ is the noise correlation function and the initial condition $c(0)=1$ (see Supplemental Material \cite{supplementary}).
Containing all the backaction effect between the probe and the noise, the convolution in Eqs. \eqref{eq:8} renders the dynamics non-Markovian. Assuming that the noise is initially in vacuum state $|\Psi_\text{E}(0)\rangle=|\{0_k\}\rangle$ and repeating the same procedure as the ideal case, we obtain
\begin{eqnarray}
\bar{M}&=&\text{Re}[e^{i\omega_{0}t}c(t)](\sinh ^{2}r-\left\vert \alpha \right\vert ^{2}),\label{eq:11}\\
\delta M&=&\{[\text{Im}(e^{i\omega _{0}t}c(t))]^{2}[|\alpha \cosh r-\alpha^{\ast }e^{i\phi }\sinh r|^{2}+\sinh ^{2}r] \nonumber\\
&&+[\text{Re}(e^{i\omega _{0}t}c(t))]^{2}[\left\vert \alpha \right\vert ^{2}+\frac{\sinh ^{2}2r}{2}]+\frac{1-\left\vert c(t)\right\vert ^{2}}{2}\nonumber\\
&&\times(\left\vert \alpha \right\vert^{2}+\sinh ^{2}r)\}^{1\over2}, \label{deltam}
\end{eqnarray}which give the distribution of the measurement results to $\hat{M}$ in the noisy situation. Then $\delta\gamma$ can be evaluated, which is analytically complicated and can be numerically solved. However, an analytical form is obtainable in the long-time limit via the asymptotic solution of Eq. \eqref{eq:8}.

The previous results are recoverable as a special case in the Markovian limit \cite{PhysRevLett.102.040403,PhysRevLett.108.130402,Lu2015}. When the probe-noise coupling is weak and the time scale of $f(t-\tau)$ is much smaller than the one of the probe, we apply the Markovian approximation to Eq. \eqref{eq:8} and obtain $c(t)=e^{-[\kappa+i(\omega_0+\gamma+\Delta)]t}$ with $\kappa=\pi J(\omega_0+\gamma)$ and $\Delta=\mathcal{P}\int_0^\infty{J(\omega)\over\omega_0+\gamma-\omega}d\omega$ \cite{PhysRevE.90.022122}. Then Eqs. \eqref{eq:11} and \eqref{deltam} leads to $\min\delta \gamma\simeq({e^{2\kappa t}-1\over2 N t^2})^{1/2}$ when $\beta=(2\sqrt{N})^{-1}$ and $ \varphi=2\phi$. Getting divergent with time, its minimum at $t=\kappa^{-1}$ returns the SNL $e\kappa(2N)^{-1/2}$. Thus, the quantum superiority of the scheme in the Markovian noise disappears completely. This is consistent to the result based on the Ramsey spectroscopy  \cite{PhysRevLett.79.3865}.

In the general non-Markovian case, the asymptotic solution of Eq. \eqref{eq:8} is obtainable by Laplace transform $\tilde{c}(p)=[p+i(\omega_{0}+\gamma)+\int_{0}^{\infty}\frac{ J(\omega)}{p+i\omega}d\omega]^{-1}$. The inverse Laplace transform to $\tilde{c}(p)$ can be done by finding its poles from
\begin{equation}\label{eq:14}
y(\varpi)\equiv\omega_{0}+\gamma-\int_{0}^{\infty}\frac{J(\omega)}{\omega-\varpi}d\omega=\varpi,~(\varpi=ip).
\end{equation}
Note that the roots $\varpi$ multiplied by $\hbar$ is the eigenenergy of the local system consisting of the probe and its environment in single-excitation subspace. To see this, we expand the eigenstate as $|\Phi\rangle=(x\hat{a}^\dag_2+\sum_ky_k\hat{b}^\dag_k)|0_2,\{0_k\}\rangle$. From the stationary Schr\"{o}dinger equation of the local system, we have $[E-\hbar(\omega_0+\gamma)]x=\sum_k\hbar g_ky_k$ and $y_k=\hbar g_k x/(E-\hbar\omega_k)$ with $E$ being its eigenenergy, which readily lead to Eq. \eqref{eq:14} with the replacement of $\varpi$ by $E/\hbar$. It implies that, although the subspaces with more excitation numbers might be involved in the dynamics, the decoherence of the probe is essentially determined by the energy-spectrum character in the single-excitation subspace. Because $y(\varpi)$ is a monotonically decreasing function with increasing $\varpi$ in the regime $\varpi<0$, Eq. \eqref{eq:14} has one isolated root $\varpi_\text{b}$ in the regime $\varpi<0$ provided $y(0)<0$. It has infinite roots in the regime $\varpi>0$, which form a continuous energy band. We call the discrete eigenstate with the isolated eigenenergy $\hbar\varpi_\text{b}$ bound state \cite{article,PhysRevA.24.2685,PhysRevA.81.052330}. Its formation has profound influences on the dissipation dynamics of the probe. This can be seen by making the inverse Laplace transform
\begin{equation}
c(t)=Ze^{-i\varpi_\text{b}t}+\int_{i\epsilon+0}^{ i\epsilon+\infty}{d\varpi\over 2\pi}\tilde{c}(-i\varpi)e^{-i\varpi t},\label{invlpc}
\end{equation}where $Z=[1+\int_{0}^\infty{J(\omega)\over (\varpi_\text{b}-\omega)^2}d\omega]^{-1}$ and the integral is from the energy band. Oscillating with time in continuously changing frequencies, the integral tends to zero in the long-time condition due to out-of-phase interference. Thus, if the bound state is absent, then $\lim_{t\rightarrow \infty}c(t)=0$ characterizes a complete decoherence, while if the bound state is formed, then $\lim_{t\rightarrow \infty}c(t)=Ze^{-i\varpi_\text{b}t}$ implies a dissipation suppression. For the Ohmic-type spectral density, it can be evaluated that the bound state is formed if $\omega_0+\gamma-\eta\omega_c\Gamma(s)\leq0$. Here $\Gamma(s)$ the Euler's $\Gamma$ function. The dominate role of the bound state played in noncanonical thermalization \cite{PhysRevE.90.022122} and quantum-correlation preservation \cite{PhysRevA.88.012129} has been revealed.

Focusing on the case in the presence of the bound state and substituting the asymptotic solution $Ze^{-i\varpi_\text{b}t}$ into Eqs. \eqref{eq:11} and \eqref{deltam}, we obtain (see Supplemental Material \cite{supplementary})
\begin{equation}
\min\delta\gamma|_{\beta=(2\sqrt{N})^{-1}}={(tN^{3/4})^{-1}\over Z}[1+{1-Z^2\over 2Z^2}N^{1\over 2}]^{1\over2},\label{PDLT}
\end{equation}when $t={(2m+1)\pi \over2|\omega_0-\varpi_\text{b}|}$ and $\varphi=2\phi$. It reduces to the ZL \eqref{znl} in the ideal case, where $Z=1$ and $\varpi_\text{b}=\omega_0+\gamma$. Equation \eqref{PDLT} remarkably reveals that, even in the long-time condition, $\delta\gamma$ asymptotically tends to the ideal ZL with $Z$ approaching 1, which is controllable by manipulating the spectral density $J(\omega)$ and the working frequency $\omega_0$ of the probe. It is in sharp contrast to the phenomenological \cite{PhysRevA.81.033819,PhysRevA.90.033846,PhysRevLett.102.040403,PhysRevA.81.033819,PhysRevA.80.013825,PhysRevA.95.053837,PhysRevA.78.063828,Gilbert:08,PhysRevA.75.053805,ZhangWP} and the Markovian approximate one \cite{PhysRevLett.102.040403,PhysRevLett.108.130402,Lu2015}, where $\delta\gamma$ gets divergent with time increasing. Indicating the significant role of the non-Markovian effect and the energy-spectrum character of the local system in the noise mechanism of quantum metrology, our result supplies a guideline to retrieve the ideal limit in the noise case by engineering the formation of the bound state.

\begin{figure}
	\centering
	\includegraphics[width=1.0\columnwidth]{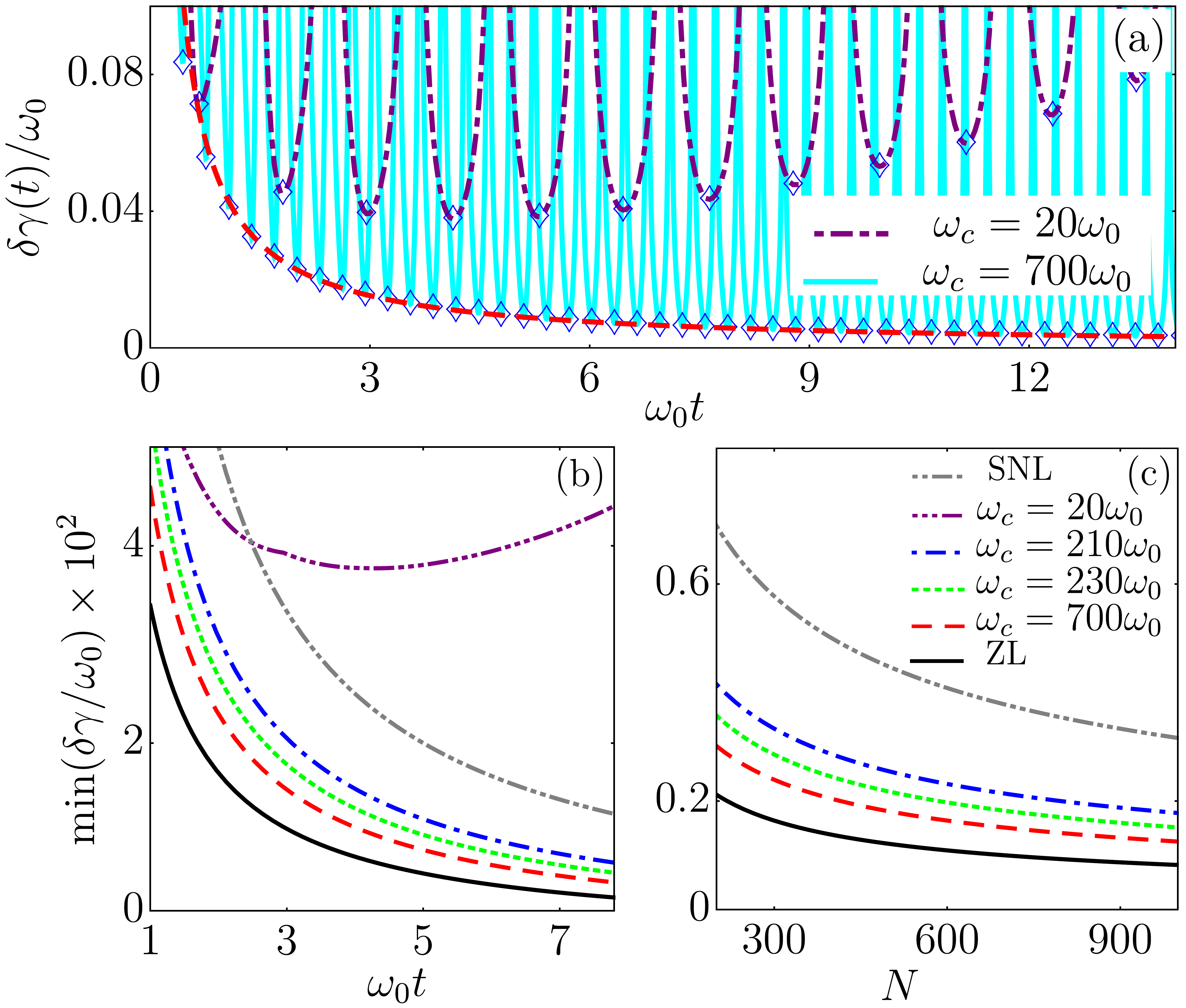}\\
	\caption {(a) Evolution of $\delta\gamma(t)$ in the absence of (purple dotdashed line) and in the presence of (cyan solid line) the bound state, where the local minima match with the curve (red dashed line) of Eq. \eqref{PDLT} in long-time condition. Dependence of the local minima of $\delta\gamma(t)$ on time (b) and $N$ (c) in different $\omega_c$. Parameters are the same as Fig. \ref{MZ} except $\beta=(2\sqrt{N})^{-1}$, $N=100$ in (b), and $t=10\omega_0^{-1}$ in (c). }
	\label{rt}
\end{figure}
	
\section{Numerical results}
To verify the distinguished role played by the bound state in the dissipation dynamics of the probe, we plot in Fig. \ref{MZ}(b) the long-time behavior of the $|c(t)|$ by numerically solving Eq. \eqref{eq:8} for the Ohmic spectral density. We can obtain from our analysis that the bound state is formed when $\omega_0+\gamma-\eta\omega_c\leq0$ for $s=1$. We really observe in Fig. \ref{MZ}(b) an abrupt change of $|c(\infty)|$ from zero to finite values exactly coinciding with $Z$ in Eq. \eqref{invlpc} with increasing $\omega_c$. Figure \ref{MZ}(c) reveals that the regime where $|c(\infty)|$ takes finite values matches with the one where a bound state is formed in the energy spectrum. It is physically understandable from the fact that the bound state, as a stationary state of the local system, would preserve the quantum coherence in its superposed components during time evolution. It is also interesting to see that quite a large value of $|c(\infty)|=Z$ approaching unity can be achieved with increasing $\omega_c$. It is readily expected from Eq. \eqref{PDLT} that the ZL is retrievable.

\begin{figure}
	\centering
	\includegraphics[width=1.0\columnwidth]{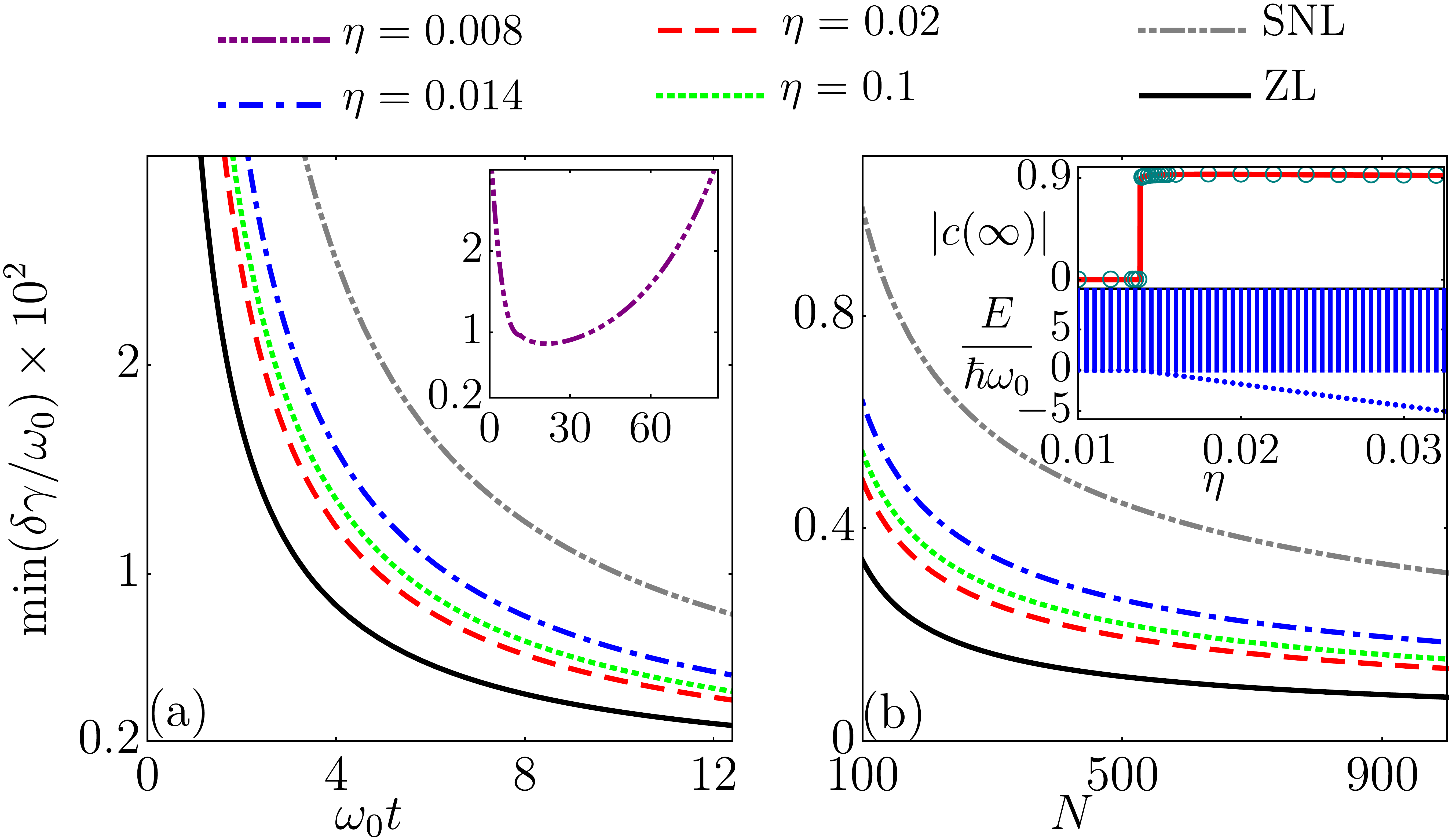}\\
	\caption {Dependence of min$(\delta\gamma/\omega_{0})$ on time (a) and $N$ (b) in difference $\eta$. The inset of (b) shows $Z$ (red solid line) coinciding with $|c(\infty)|$ (dark cyan circles) and the energy spectrum. Parameters are the same as Fig. \ref{rt} except $\omega_c=300\omega_0$. }
	\label{fig3}
\end{figure}

With the numerical result of Eq.~\eqref{eq:8} at hands, we can calculate the exact precision $\delta \gamma(t)$ from Eqs. \eqref{eq:11} and \eqref{deltam}. Figure \ref{rt}(a) shows the evolution of $\delta\gamma(t)$ in different $\omega_c$. Oscillating with time, $\delta\gamma(t)$ takes its best values at the local minima marked by the blue diamonds. It is seen that the local minima become larger and larger with time when the bound state is absent, which is consistent with the Markovian result. The noisy metrology scheme performs worse and worse with increasing the encoding time in this situation. However, as long as the bound state is formed, the profile of the local minima gets to be a decreasing function with time. Thus, the bound state makes the superiority of the encoding time as a resource in the ideal metrology case recovered. The red dashed line in Fig. \ref{rt}(a) gives $\min(\delta\gamma)$ evaluated from Eq. \eqref{PDLT}, which matches with the local minima of $\delta\gamma(t)$ in the long-time condition. This verifies the validity of our result \eqref{PDLT}. Focusing on the case in the presence of the bound state, we plot in Fig. \ref{rt}(b) $\min(\delta\gamma)$ in different $\omega_c$. It reveals that, with the formation of the bound state, not only the SNL can be surpassed, but also the ideal ZL is asymptotically retrieved. This is further verified by $\min(\delta\gamma)$ as function of total photon number [see Fig. \ref{rt}(c)]. Once again, it validates the scaling \eqref{PDLT}. Furthermore, with the increasing of $Z$ accompanying the increasing of $\omega_c$, $\min(\delta\gamma)$ gets nearer and nearer the ZL. All the results confirm our expectation that the precision asymptotically matching the analytical scaling \eqref{PDLT} approaches the ZL with the formation of the bound state. Besides tuning $\omega_c$, the bound-state-favored retrieving of the ZL is also obtained by tuning $\eta$. Figure \ref{fig3} gives the dependence of $\min\delta\gamma$ on time and $N$ in different $\eta$. It verifies again our conclusion that the ZL is asymptotically recoverable when the bound state is formed.

Note that our finding on the bound-state-favored retrieving of the ideal precision in noisy situation is readily applicable to the optimal measurement protocol \cite{PhysRevLett.100.073601} by calculating quantum Fisher information, where the Heisenberg-limit recovery is expected. Further, our result in Eq. \eqref{PDLT} is independent on the form of the spectral density. Although only the Ohmic form is considered, our result can be generalized to other cases, where the specific condition to support the bound state might be quantitatively different, but the conclusion on the bound-state-favored retrieving of the ideal precision is the same. With the rapid development of reservoir engineering technique \cite{ER1,Kienzler53}, people have obtained rich ways to engineer the spectral densities. The non-Markovian effect has been observed in the linear optical systems \cite{Liu2011,Benardes2015}, on which the MZI is based. A sub-Ohmic spectrum has been engineered and the non-Markovian effect is observed in a micromechanical system \cite{Gro2015}, which shares the similar characters with the probe in our system. An all-optical non-Markovian quantum simulator has been proposed \cite{PhysRevA.91.012122}. The Ohmic spectral density is possible to be controlled in trapped ion system \cite{PhysRevA.78.010101}.
The bound state has been observed in circuit QED \cite{Liu2016} and ultracold atom \cite{Kri2018} systems. All these progresses show a crucial support to our argument and indicate that our finding is realizable in the state-of-art technique of quantum-optics experiments.

\section{Conclusions}\label{conclusion}
In summary, we have microscopically studied the non-Markovian noise effect on the MZI-based quantum metrology scheme. An exact scaling relation of the precision to the photon number is derived in the long-encoding-time condition. It is remarkable to find that the ideal ZL is asymptotically recoverable. We reveal that the non-Markovian effect and the formation of a bound state between the quantum probe and its environment are two essential reasons for retrieving the ZL: The bound state supplies the intrinsic ability and the non-Markovian effect supplies the dynamical way. Our result suggests a guideline to experimentation to implement the ultrasensitive measurement in the practical noise situation by engineering the formation of the bound state.
\section{Acknowledgments}
The work is supported by the National Natural Science Foundation (Grant Nos. 11875150, 11474139, 11674139, and 11834005) and by the Fundamental Research Funds for the Central Universities of China.

\bibliography{reference}

\begin{thebibliography}{78}
\bibitem [{HLL()}]{HLL}%
  \BibitemOpen
  \href@noop {} {\ \bibinfo {series} {It was proven that the optimal limit
  using squeezed state is the Heisenberg limit $N^{-1}$ with a full knowledge
  on the fluctuation of the photon number in the two output ports when the
  average photon number from the squeezed state is equal to the one of the
  coherent light \cite{PhysRevLett.100.073601}.}}\BibitemShut {Stop}%
\end{thebibliography}

\end{document}


\title{Supplemental material for ``Retrieving ideal precision in noisy quantum optical metrology''}
\author{Kai Bai}
\affiliation{School of Physical Science and Technology \& Key Laboratory for Magnetism and Magnetic Materials of the MoE, Lanzhou University, Lanzhou 730000, China}
\author{Zhen Peng}
\affiliation{School of Physical Science and Technology \& Key Laboratory for Magnetism and Magnetic Materials of the MoE, Lanzhou University, Lanzhou 730000, China}
\author{Hong-Gang Luo}
\affiliation{School of Physical Science and Technology \& Key Laboratory for Magnetism and Magnetic Materials of the MoE, Lanzhou University, Lanzhou 730000, China}
\affiliation{Beijing Computational Science Research Center, Beijing 100084, China}
\author{Jun-Hong An}
\email{anjhong@lzu.edu.cn}
\affiliation{School of Physical Science and Technology \& Key Laboratory for Magnetism and Magnetic Materials of the MoE, Lanzhou University, Lanzhou 730000, China}

\maketitle
\section{Decoherence dynamics}

In this section, we give the detailed derivation of the dynamical equations of motion of the noisy quantum metrology. The total Hamiltonian governing the parameter encoding reads
 \begin{equation}
 \hat{H}=\hat{H}_0+\hbar\sum_k\omega_k[\hat{b}_k^\dag\hat{b}_k+g_k(\hat{a}_2\hat{b}_k^\dag+\text{H.c.})],
 \end{equation}
with $\hat{H}_0=\hbar\omega_0\sum_{m=1,2}\hat{a}_m^\dag\hat{a}_m+\hbar\gamma\hat{a}_2^\dag\hat{a}_2$. In the Heisenberg picture with $\hat{\bullet}(t)=\hat{U}^\dag(\gamma,t)\hat{\bullet}\hat{U}(\gamma,t)$ and $\hat{U}(\gamma,t)=\exp({-i\hat{H}t/\hbar})$, the equations of motion of the field operator read
\begin{eqnarray}
\dot{\hat{a}}_1(t)+i\omega_0\hat{a}_1(t)&=&0,\label{eq:2}\\
\dot{\hat{a}}_{2}(t)+i(\gamma+\omega_{0})\hat{a}_{2}(t)+i\sum_{k}g_{k}\hat{b}_{k}(t)&=&0,\label{eq:3}\\
\dot{\hat{b}}_{k}(t)+i\omega_{k}\hat{b}_{k}(t)+ig_{k}\hat{a}_{2}(t)&=&0.\label{eq:4}
\end{eqnarray}
Substituting the formal solution of Eq. (\ref{eq:4})
\begin{equation}
\hat{b}_k(t)=\hat{b}_{k}e^{-i\omega _{k}t}-ig_{k}\int_{0}^{t}e^{-i\omega _{k}(t-\tau
)}\hat{a}_{2}(\tau )d\tau
\end{equation}into Eq. (\ref{eq:3}), we obtain
\begin{eqnarray}
\dot{\hat{a}}_{2}(t)&=&-i(\gamma+\omega_{0})\hat{a}_{2}(t)-\int_{0}^{t}f(t-\tau)\hat{a}_{2}(\tau)d\tau \nonumber\\
&&-i\sum_{k}g_{k}\hat{b}_{k}e^{-i\omega_{k}t},\label{eq:5}
\end{eqnarray}
where $f(t-\tau)=\int_0^\infty J(\omega)e^{-i\omega(t-\tau)}$ is the noise correlation function. The linearity of Eq.~\eqref{eq:3} implies that $\hat{a}_{2}(t)$ can be expanded as \cite{PhysRevA.81.052105}
\begin{equation}
\hat{a}_{2}(t)=c(t)\hat{a}_{2}+\sum_{k}d_{k}(t)\hat{b}_{k}.\label{exm}
\end{equation} One can check $|c(t)|^{2}+\sum_k|d_{k}(t)|^{2}=1$ from $[a_{2}(t),a_{2}^{\dagger}(t)]=1$. Substituting this expansion into Eq. (\ref{eq:5}), we obtain
\begin{eqnarray}\label{eq:6}
\dot{c}(t)+i(\gamma+\omega_{0})c(t)+\int_{0}^{t}f(t-\tau)c(\tau)d\tau=0\label{integrd}
\end{eqnarray}
under the initial conditions $c(0)=1$. Containing all the backaction effect between the probe and the noise, the convolution in Eqs. \eqref{eq:6} renders the dynamics non-Markovian.

In the special case of weak coupling, we can make Markovian approximation to Eq. \eqref{eq:6}. Defining $c(t)=e^{-i(\omega_0+\gamma)t}c'(t)$, we can rewrite Eq. \eqref{eq:6} as
\begin{equation}
\dot{c}'(t)+\int_0^\infty d\omega J(\omega)\int_0^td\tau c'(\tau)e^{-i(\omega-\omega_0-\gamma)(t-\tau)}=0.\label{mkv1}
\end{equation}
Then, we take the Markovian approximation $c'(\tau)\simeq c'(t)$, namely, neglecting any memory effect regarding the earlier time. The approximation is valid when the correlation time of the environment is much smaller then the typical time scale of system evolution. Also under this assumption we can extend the upper limit of the $\tau$ integration in Eq. \eqref{mkv1} to infinity and use the equality
\begin{equation}
\lim_{t\rightarrow\infty}\int_0^t d\tau e^{-ix(t-\tau)}=\pi\delta(x)-i\mathcal{P}\left({1\over x}\right)
\end{equation}
where $\mathcal{P}$ and the delta function denote the Cauchy principal value and the singularity, respectively. Equation \eqref{mkv1} is reduced to a linear ordinary differential equation. The solution of $c(t)$ can then be easily obtained as
\begin{equation}c(t)=e^{-[\kappa+i(\omega_0+\gamma+\Delta)]t} \label{MKL}
\end{equation} with $\kappa=\pi J(\omega_0+\gamma)$ and $\Delta=\mathcal{P}\int_0^\infty d\omega{J(\omega)\over \omega_0+\gamma-\omega}$. Here $\kappa$ is the constant loss rate of photon and $\Delta$ is the frequency shift induced by the environment  \cite{PhysRevE.90.022122}.

\begin{figure}
	\centering
	\includegraphics[width=.5\columnwidth]{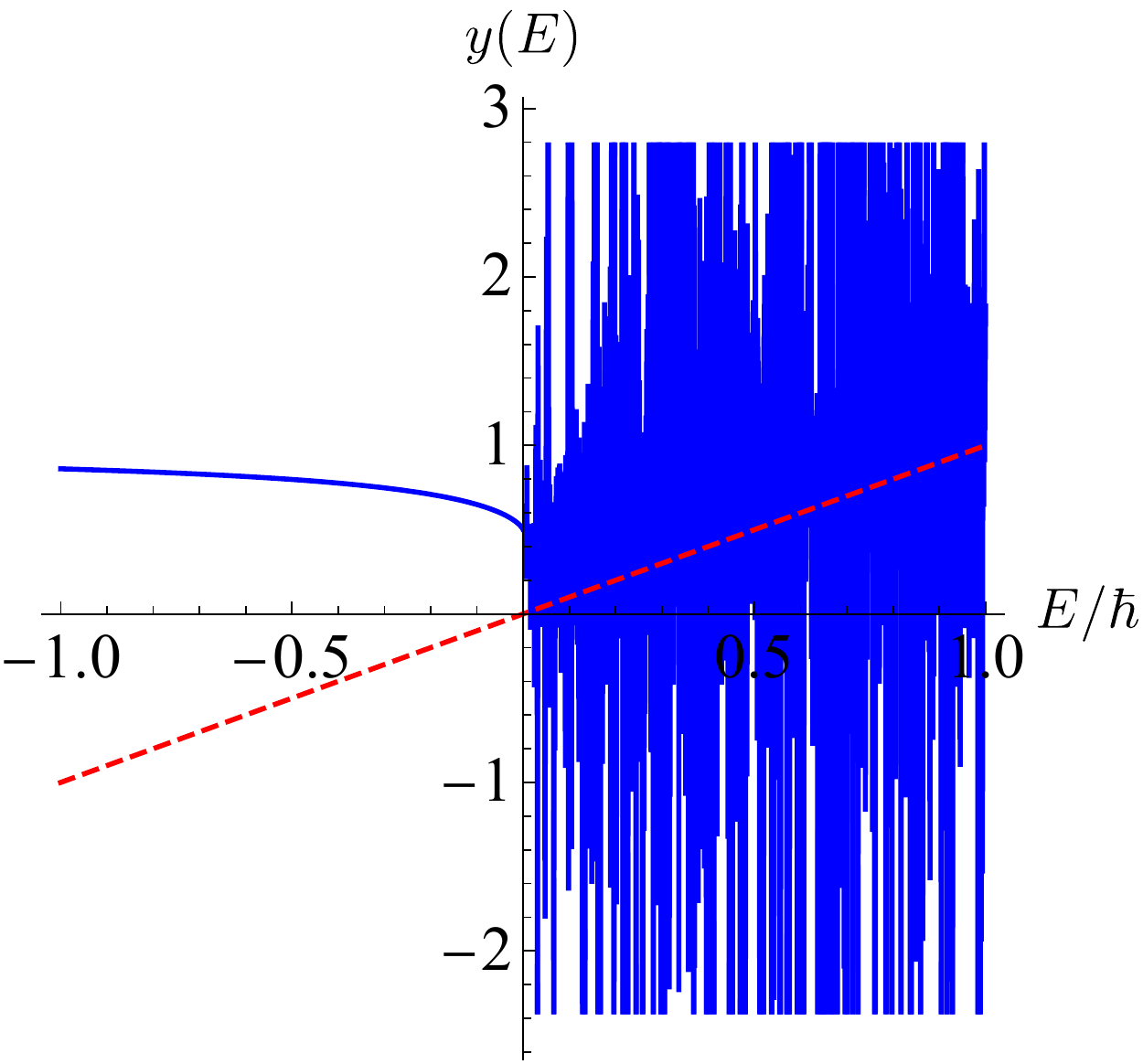}~\includegraphics[width=.5\columnwidth]{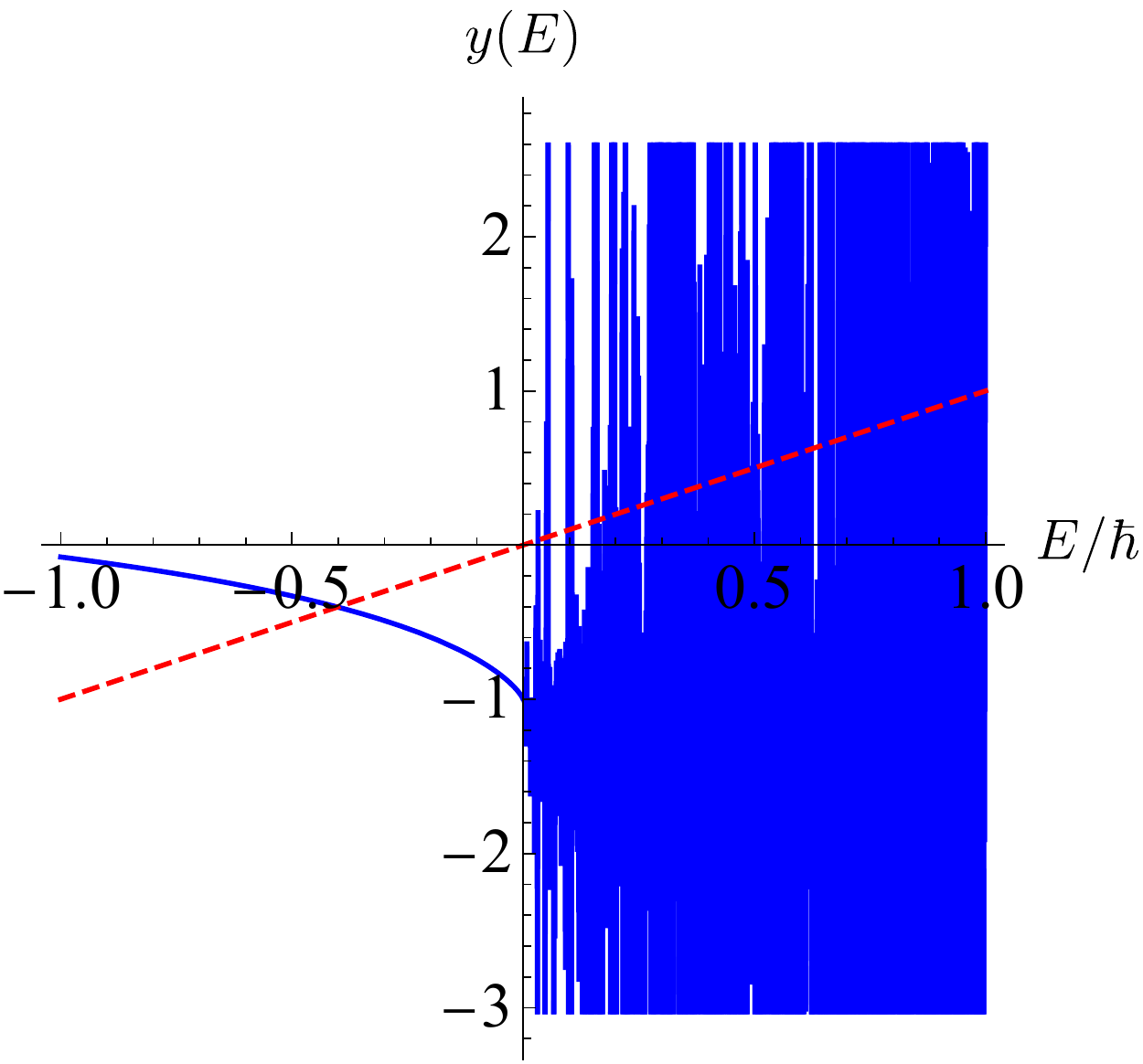}\\
	\caption {Solution of Eq. \eqref{yee} determined by the intersectors of two curves when $y(0)>0$ in (a) and $y(0)<0$ in (b). The parameters are the Ohmic spectral density with $\eta=1$ and $\omega_c=0.5(\omega_0+\gamma)$ in (a) and $2(\omega_0+\gamma)$ in (b).  }
	\label{fs1}
\end{figure}

In the general non-Markovian case, the solution of Eq. (\ref{eq:6}) can be analyzed by Laplace transform $\tilde{c}(p)=\int_0^\infty c(t)e^{-pt}dt$ as \cite{PhysRevLett.109.170402}
\begin{equation}
\tilde{c}(p)=[p+i(\omega_0+\gamma)+\tilde{f}(p)]^{-1} \label{cs}
\end{equation}with $\tilde{f}(p)=\int_0^\infty {J(\omega)\over p+i\omega}d\omega$ being the Laplace transform of the noise correlation function $f(t)$. The form of $c(t)$ is obtained by the inverse Laplace transform to $\tilde{c}(p)$, which can be performed by finding the pole of Eq. (\ref{cs}) from
\begin{equation}
y(\varpi)\equiv\omega_{0}+\gamma-\int_{0}^{\infty}\frac{J(\omega)}{\omega-\varpi}d\omega=\varpi,~(\varpi=ip). \label{yee}
\end{equation}
As demonstrated in the main text, the solutions of Eq. \eqref{yee} multiplied by $\hbar$ are just the eigen energies of the local system composed of the second optical field and its environment. From the analytical property of $y(\varpi)$, we conclude that Eq. \eqref{yee} has infinite solutions in the regime $\varpi>0$, which after multiplied by $\hbar$ form a continuous energy band. As long as $y(0)<0$, Eq. \eqref{yee} has one and only one solution in regime $\varpi<0$ out of the formed continuous energy band [see Fig. \ref{fs1}]. We call the eigestate corresponding to such isolated eigenenergy bound state. Using the residue theorem, the inverse Laplace transform can be calculated as
\begin{equation}
c(t)=Ze^{-i\varpi_\text{b}t}+\int_{i\epsilon+0}^{ i\epsilon+\infty}{d\varpi\over 2\pi}\tilde{c}(-i\varpi)e^{-i\varpi t},
\end{equation}where the first term with $Z=[1+\int_{0}^\infty{J(\omega)\over (\varpi_\text{b}-\omega)^2}d\omega]^{-1}$ is from the formed bound state and the second term is from the continuous-band states. Due to the destructive interference among the components with different $\varpi$, the second term tends to vanish in the long-time limit. Therefore, we have
\begin{equation}
\lim_{t\rightarrow\infty}c(t)=\left\{\begin{aligned}
                                &0,\hspace{1.05cm} y(0)>0 \\
                                &Ze^{-i\varpi_\text{b}t}, y(0)<0
                              \end{aligned}\right..\label{stda}
\end{equation}
Such long-time behavior is verified by numerically calculating Eq. \eqref{integrd} [see Fig. 1(b) in the main text].

\section{Metrology precision in noisy case}
In this section, we give the detailed derivation of metrology precision. Assuming that the noise is initially in vacuum state $|\Psi_\text{E}(0)\rangle=|\{0_k\}\rangle$, we have
\begin{eqnarray}
\bar{M}=\langle \Psi_\text{in},\{0_{k}\}|\hat{V}^{\dag }\hat{U}^{\dag }(\gamma,t)\hat{V}^{\dag
}\hat{M}\hat{V}\hat{U}(\gamma,t)\hat{V}|\Psi_\text{in},\{0_{k}\}\rangle,~~~~~~\label{def}
\end{eqnarray}
where $\hat{V}=\exp[i\frac{\pi}{4}(\hat{a}_{1}^{\dagger}\hat{a}_{2}+\hat{a}_{2}^{\dagger}\hat{a}_{1})]$, $\hat{D}_{\hat{a}}=\exp(\alpha \hat{a}^{\dagger}-\alpha^{*}\hat{a})$ with $\alpha=|\alpha|e^{i\varphi}$, $\hat{S}_{\hat{a}}=\exp[\frac{1}{2}(\xi^{*}\hat{a}^{2}-\xi \hat{a}^{\dagger2})]$ with $\xi=re^{i\phi}$, and $\hat{M}=\hat{a}^\dag_1\hat{a}_1-\hat{a}^\dag_2\hat{a}_2$. Because $\hat{U}^\dag(\gamma,t)\hat{a}_1\hat{U}(\gamma,t)=\hat{a}_1e^{-i\omega_0t}$ in Eq. \eqref{eq:2} and $\hat{U}^\dag(\gamma,t)\hat{a}_2\hat{U}(\gamma,t)=c(t)\hat{a}_{2}+\sum_{k}d_{k}(t)\hat{b}_{k}$ in Eq. \eqref{exm}, Eq. \eqref{def} can be evaluated as
\begin{equation}
\bar{M}=\text{Re}[e^{i\omega_{0}t}c(t)](\sinh ^{2}r-\left\vert \alpha \right\vert ^{2}).\label{eq:8}
\end{equation}In the same manner, we also obtain
\begin{eqnarray}
\overline{M^{2}}
&=&[\text{Im}(e^{i\omega _{0}t}c(t))]^{2}[|\alpha \cosh r-\alpha ^{\ast
}e^{i\phi }\sinh r|^{2}+\sinh ^{2}r]\nonumber \\
&&+[\text{Re}(e^{i\omega _{0}t}c(t))]^{2}[\left\vert \alpha \right\vert
^{2}+(\left\vert \alpha \right\vert ^{2}-\sinh ^{2}r)^{2}+2\sinh ^{2}r \nonumber \\ && \times\cosh
^{2}r] +\frac{1-\left\vert c(t)\right\vert ^{2}}{2}(\left\vert \alpha \right\vert
^{2}+\sinh ^{2}r)
\end{eqnarray}Then the variance of $\hat{M}$ can be calculated as
\begin{eqnarray}
\delta M&=&\{[\text{Im}(e^{i\omega _{0}t}c(t))]^{2}[|\alpha \cosh r-\alpha^{\ast }e^{i\phi }\sinh r|^{2}+\sinh ^{2}r] \nonumber\\
&&+[\text{Re}(e^{i\omega _{0}t}c(t))]^{2}[\left\vert \alpha \right\vert ^{2}+\frac{\sinh ^{2}2r}{2}]+\frac{1-\left\vert c(t)\right\vert ^{2}}{2}\nonumber\\
&&\times(\left\vert \alpha \right\vert^{2}+\sinh ^{2}r)\}^{1\over2}, \label{deltam}
\end{eqnarray}

First, in the special case of Markovian limit, substituting Eq. \eqref{MKL} into Eq. \eqref{deltam} and using $\delta \gamma=\frac{\delta M}{|\partial\bar{M}/\partial\gamma|}$, we have
\begin{equation}
\min \delta\gamma={[(1-\beta)e^{-2r}+\beta+{e^{2\kappa t}-1\over 2}]^{1\over 2}\over t\sqrt{N}|1-2\beta|}
\end{equation}
when $\phi=2\varphi$ and $\gamma t=(2m+1)\pi/2$ with $m\in \mathbb{Z}$. Here the unimportant frequency shift $\Delta$ has been neglected. Substituting $e^{-2r}\simeq {1\over 4N\beta}$ in the large $N$ case and optimizing $\beta$, we obtain
 \begin{equation}
\min \delta\gamma|_{\beta=(2\sqrt{N})^{-1}}=\Big({N^{-1\over 2}+{e^{2\kappa t}-1\over 2}\over N t^2}\Big)^{1\over 2}\simeq  \Big({e^{2\kappa t}-1\over 2N t^2}\Big)^{1\over 2}.
\end{equation}
It indicates that, after decreasing with $t$ in a short-time scale, $\min \delta\gamma|_{\beta=(2\sqrt{N})^{-1}}$ gets larger and larger with $t$. It means that the superiority of the encoding time as a resource in the ideal metrology complectly disappears in the Markovian noise. Optimizing $t$, we can obtain the best precision
\begin{equation}
\min \delta\gamma|_{\beta=(2\sqrt{N})^{-1},t=\kappa^{-1}}=e\kappa (2\sqrt{N})^{-1}.
\end{equation} This scaling relation with $N$ is just the shot-noise limit. Thus the Markovian noise forces the metrology precision back to the classical limit. This is consistent to the noisy metrology result based on the Ramsey spectroscopy  \cite{PhysRevLett.79.3865}.

Second, in the general non-Markovian case, $\delta\gamma$ can be evaluated from Eqs. \eqref{deltam} and \eqref{eq:8}. It is analytically complicated. With $c(t)$ numerically calculated from Eq. \eqref{eq:6}, the exact behavior of $\delta\gamma$ can be obtained [see Fig. 2(a) in the main text]. Focusing on the case in the presence of the bound state, we also can obtain the analytical result of $\delta\gamma$ in the long-encoding-time limit. Using the long-time result \eqref{stda}, we can calculate
\begin{equation}
\min \delta\gamma={[(1-\beta)e^{-2r}+\beta+{1-Z^2\over 2Z^2}]^{1\over 2}\over Z t\sqrt{N}|1-2\beta|}
\end{equation}
when $\phi=2\varphi$ and $\gamma t=(2m+1)\pi/2$ with $m\in \mathbb{Z}$. Its optimal value with respect to $\beta$ reads
\begin{equation}
\min \delta\gamma|_{\beta=(2\sqrt{N})^{-1}}={(tN^{3/4})^{-1}\over Z}[1+{1-Z^2\over 2Z^2}N^{1\over 2}]^{1\over2}. \label{opt}
\end{equation}This is the exact scaling relation of the metrology precision in the long-encoding-time condition. The precision evaluated via Eq. \eqref{opt} matches well with the local minima obtained via the numerically solving Eq. \eqref{eq:6} in the long-encoding-time limit [see Fig. 2(a) in the main text].

\bibliography{reference}